\documentclass{article}
\usepackage{amsmath, amssymb, amsfonts}
\title{ On a particular Morris - Thorne wormhole } 
\author{Hristu Culetu, \\Ovidius University, Dept.of Physics and Electronics, \\B-dul Mamaia 124, 900527 Constanta, Romania, \\e-mail : hculetu@yahoo.com}

\begin{document}
\numberwithin{equation}{section}
\pagenumbering{arabic}
\maketitle
\newcommand{\fv}{\boldsymbol{f}}
\newcommand{\tv}{\boldsymbol{t}}
\newcommand{\gv}{\boldsymbol{g}}
\newcommand{\OV}{\boldsymbol{O}}
\newcommand{\wv}{\boldsymbol{w}}
\newcommand{\WV}{\boldsymbol{W}}
\newcommand{\NV}{\boldsymbol{N}}
\newcommand{\hv}{\boldsymbol{h}}
\newcommand{\yv}{\boldsymbol{y}}
\newcommand{\RE}{\textrm{Re}}
\newcommand{\IM}{\textrm{Im}}
\newcommand{\rot}{\textrm{rot}}
\newcommand{\dv}{\boldsymbol{d}}
\newcommand{\grad}{\textrm{grad}}
\newcommand{\Tr}{\textrm{Tr}}
\newcommand{\ua}{\uparrow}
\newcommand{\da}{\downarrow}
\newcommand{\ct}{\textrm{const}}
\newcommand{\xv}{\boldsymbol{x}}
\newcommand{\mv}{\boldsymbol{m}}
\newcommand{\rv}{\boldsymbol{r}}
\newcommand{\kv}{\boldsymbol{k}}
\newcommand{\VE}{\boldsymbol{V}}
\newcommand{\sv}{\boldsymbol{s}}
\newcommand{\RV}{\boldsymbol{R}}
\newcommand{\pv}{\boldsymbol{p}}
\newcommand{\PV}{\boldsymbol{P}}
\newcommand{\EV}{\boldsymbol{E}}
\newcommand{\DV}{\boldsymbol{D}}
\newcommand{\BV}{\boldsymbol{B}}
\newcommand{\HV}{\boldsymbol{H}}
\newcommand{\MV}{\boldsymbol{M}}
\newcommand{\be}{\begin{equation}}
\newcommand{\ee}{\end{equation}}
\newcommand{\ba}{\begin{eqnarray}}
\newcommand{\ea}{\end{eqnarray}}
\newcommand{\bq}{\begin{eqnarray*}}
\newcommand{\eq}{\end{eqnarray*}}
\newcommand{\pa}{\partial}
\newcommand{\f}{\frac}
\newcommand{\FV}{\boldsymbol{F}}
\newcommand{\ve}{\boldsymbol{v}}
\newcommand{\AV}{\boldsymbol{A}}
\newcommand{\jv}{\boldsymbol{j}}
\newcommand{\LV}{\boldsymbol{L}}
\newcommand{\SV}{\boldsymbol{S}}
\newcommand{\av}{\boldsymbol{a}}
\newcommand{\qv}{\boldsymbol{q}}
\newcommand{\QV}{\boldsymbol{Q}}
\newcommand{\ev}{\boldsymbol{e}}
\newcommand{\uv}{\boldsymbol{u}}
\newcommand{\KV}{\boldsymbol{K}}
\newcommand{\ro}{\boldsymbol{\rho}}
\newcommand{\si}{\boldsymbol{\sigma}}
\newcommand{\thv}{\boldsymbol{\theta}}
\newcommand{\bv}{\boldsymbol{b}}
\newcommand{\JV}{\boldsymbol{J}}
\newcommand{\nv}{\boldsymbol{n}}
\newcommand{\lv}{\boldsymbol{l}}
\newcommand{\om}{\boldsymbol{\omega}}
\newcommand{\Om}{\boldsymbol{\Omega}}
\newcommand{\Piv}{\boldsymbol{\Pi}}
\newcommand{\UV}{\boldsymbol{U}}
\newcommand{\iv}{\boldsymbol{i}}
\newcommand{\nuv}{\boldsymbol{\nu}}
\newcommand{\muv}{\boldsymbol{\mu}}
\newcommand{\lm}{\boldsymbol{\lambda}}
\newcommand{\Lm}{\boldsymbol{\Lambda}}
\newcommand{\opsi}{\overline{\psi}}
\renewcommand{\tan}{\textrm{tg}}
\renewcommand{\cot}{\textrm{ctg}}
\renewcommand{\sinh}{\textrm{sh}}
\renewcommand{\cosh}{\textrm{ch}}
\renewcommand{\tanh}{\textrm{th}}
\renewcommand{\coth}{\textrm{cth}}

\begin{abstract}
The properties of a particular Misner - Thorne wormhole are investigated. The ''exotic stress-energy'' needed to maintain the wormhole open corresponds to a massless scalar field whose Lagrangean density contains a negative kinetic term. While the Komar energy of the spacetime is vanishing due to the negative energy density and radial pressure, the ADM energy is (minus) the Planck energy. The timelike geodesics are hyperbolae and any static observer is inertial. The null radial trajectories are also hyperbolae and Lorentz invariant as Coleman- de Luccia expanding bubble or Ipser-Sikivie domain wall. Using a different equation of state for the fluid on the dynamic wormhole throat of Redmount and Suen, we reached an equation of motion for the throat (a hyperbola) that leads to a negative surface energy density and the throat expands with the same acceleration $2\pi |\sigma|$ as the Ipser-Sikivie domain wall.
 \end{abstract}
 
\section{Introduction}

Wormholes (WHs) are defined as topological structures connecting two asymptotically flat spacetimes. It is well-known that Lorenzian WHs are threaded by matter that violates the energy conditions (one needs ''exotic matter'' to avoid the WH collapse) \cite{HE, MT, EH, RS, AP, MV, HMV, ES, LL, SK, MH}. The cut-and-paste prescription may be used for to build static and dynamic WHs by surgically - grafting together two asimptotically flat spacetimes \cite{MV, LL}. 

 The surgery procedure leads to a nonzero energy-momentum tensor on the boundary layer between the two regions. As Morris and Thorne \cite{MT} have shown, to make the WH traversable we must enforce the presence of ''exotic stress-energy'' on its throat which asks for ''the flare-out condition'' \cite{ES, SK} to be satisfied. Ellis \cite{HE} introduced a ''drainhole'' (a topological hole) in the Schwarzschild manifold to avoid the singularity at the origin. His geometry is coupled to a scalar field with a reversed sign kinetic term, to have geodesic completeness.
 
 A significant contribution to the WH physics was brought by the above-mentioned paper by Morris and Thorne \cite{MT} and by Morris, Thorne and Yurtsever \cite{MTY}. They showed that the Schwarzschild WH is not traversable and constructed traversable WHs which possess matter on their throat, violating all the energy conditions. 
 
 Using the ''junction formalism'' Visser \cite{MV} assumed that all ''exotic matter'' is confined to a thin boundary layer between asimptotically flat universes. He applied the formalism both for static and dynamic WHs. While Harris \cite{EH} has shown that a scalar field with negative energy is necessary to maintain his WH connecting two Reisnner - Nordstrom universes, Redmount and Suen \cite{RS} consider a simple WH geometry as a model for topological fluctuations in the Planck scale spacetime foam. They proved that the WH is quantum-mechanically unstable and its wave function has to leak to macroscopic radii. 
 
 Plane symmetric thin shell WHs have been analyzed by Lemos and Lobo \cite{LL}. They found that thin shell WHs made of a dark energy fluid or of a cosmological constant fluid are stable but those made of phantom energy are unstable. More recently, Kim \cite{SK} re-considered the flare-out condition (the minimality of the WH throat), checked the finiteness of the traversal pressures and studied their physical meaning.  
 
 We investigated in this paper the timelike and null geodesics for the special case of the Morris - Thorne wormhole, namely when the shape function is $B(r) = b^{2}/r$ ($b$ is here the throat radius) and for a vanishing redshift function \cite{MT, EH, SK}. We fould that the radial geodesics are hyperbolae and become straightlines when $r >> b$. They resemble the expanding Lorentz-invariant Coleman - de Luccia bubble \cite{CL} or Ipser and Sikivie planar domain wall \cite{IS} which, in Minkowski coordinates is not a plane at all but an accelerating sphere.
 
 The stress tensor of the exotic matter corresponds to that of a massless scalar field with a negative sign for the kinetic term of the Lagrangean density \cite{EH}. The Komar energy of the system is found to vanish but its matter energy \cite{SW, HC1} equals that obtained by Ellis \cite{HE} for the scalar field energy. We also computed the quasilocal Misner - Sharp energy \cite{CG1, CG2, NY, GGMP, HC1} together with its Ricci and Weyl parts. We take the WH throat radius $b$ of the order of the Planck length and conjecture that it corresponds to the minimum length for the Minkowski interval $\sqrt{r^{2} - t^{2}}$ (see also \cite{DK, KP}). In the last section we propose a different equation of state for the fluid on the boundary of the Redmount - Suen dynamic WH and show that the throat evolves hyperbolically.
  
 Throughout the paper geometrical units $G = c = \hbar = 1$ are used, unless otherwise specified.
 
 \section{Morris - Thorne traversable wormhole}
 Let us begin with the static, spherically symmetric Morris - Thorne wormhole \cite{MT}
     \begin{equation}
   ds^{2} = -e^{2 \Phi} dt^{2} + \frac{dr^{2}}{1 - \frac{B}{r}} + r^{2} d \Omega^{2} 
 \label{2.1}
 \end{equation}
 where $\Phi = \Phi(r)$ determines the gravitational redshift , $B = B(r)$ is the shape function (it gives the spatial shape of the WH when viewed in an embedding diagram) and $d \Omega^{2} = d\theta^{2} + sin^{2}\theta d\phi^{2}$ stands for the metric on the unit 2-sphere. The radial coordinate $r$ is not monotonic: it decreases from infinity to a minimum value $b$ in the lower universe and increases from $b$ to infinity as one moves into the upper universe \cite{MT}. 
 
 Our aim in this paper is to study the special case $\Phi = 0$ and $B(r) = b^{2}/r$  \cite{MT, SK} since the lack of tidal acceleration ($\Phi = 0$) allow travel through the throat \cite{BD}. Therefore, the metric (2.1) can be written as 
 \begin{equation}
   ds^{2} = - dt^{2} + \frac{dr^{2}}{1 - \frac{b^{2}}{r^{2}}} + r^{2} d \Omega^{2} 
 \label{2.2}
 \end{equation}
where $r \geq b$. The singularity at $r = 0$ is not part of the manifold and so causes no concern for our analysis \cite{BD}. Keeping in mind that the above WH will be related to the Planck scale spacetime foam, we take the constant $b$ to be of the order of the Planck length $l_{P} \approx 10^{-33} cm$ although on larger scale the spacetime appears smooth and simply connected. A microscopic WH might in principle be extracted from the foam to produce a traversable macroscopic WH \cite{MTY}. As Morris and Thorne (see also \cite{HE, EH, AP}) have shown, by means of the  change $l = \pm{\sqrt{r^{2} - b^{2}}}$ of the radial coordinate the line-element (2.2) acquires the form
 \begin{equation}
   ds^{2} = - dt^{2} + dl^{2} + (l^{2} + b^{2}) d \Omega^{2} 
 \label{2.3}
 \end{equation}
where $l$ (the proper radial coordinate) covers the entire real axis. The above spacetime, like (2.2), has two asymptotically flat regions $l \rightarrow \infty$ and $l \rightarrow -\infty$ and has no horizons.

\section{Anisotropic fluid stress tensor}
 For the special static traversable WH (2.2) to be a solution of Einstein's equations $G_{ab} = 8\pi T_{ab}~(a, b = 0, 1, 2 ,3)$, the following stress tensor is necessary on their r.h.s. \cite{MT, EH, SK}
  \begin{equation}
  8\pi T^{a}_{~b} = diag\left(\frac{b^{2}}{r^{4}}, - \frac{b^{2}}{r^{4}}, \frac{b^{2}}{r^{4}}, \frac{b^{2}}{r^{4}}\right)
 \label{3.1}
 \end{equation} 
which leads to $\rho = - b^{2}/8\pi r^{4} = p_{r} = - p_{\theta} = - p_{\phi}$. $\rho$ is the energy density of the fluid, $p_{r}$ is the radial pressure and $p_{\theta},~p_{\phi}$ are the ''angular'' pressures. We observe that the energy density and the radial pressure are negative (as for exotic matter). We suppose $T_{ab}$ is spread all over the spacetime and its components reach the extremal values close to the throat $r = b$. The stress tensor corresponds to an anisotropic fluid because of the nonequal principal pressures. It can be written in the general form
   \begin{equation}
   T^{a}_{~b} = (p + \rho)u^{a}u_{b} + p\delta^{a}_{b} +\pi^{a}_{~b}
 \label{3.2}
 \end{equation} 
where $p = (1/3)T^{i}_{i} = b^{2}/3r^{4}$ is the average pressure ($i = 1, 2, 3), u^{a} = (1, 0, 0, 0)$ is the velocity vector field of a static observer and 
   \begin{equation}
  \pi ^{a}_{~b} = diag\left(0, - \frac{b^{2}}{6\pi r^{4}}, \frac{b^{2}}{12\pi r^{4}}, \frac{b^{2}}{12\pi r^{4}}\right)  
 \label{3.3}
 \end{equation} 
is the anisotropic stress tensor with $\pi^{a}_{~a} = 0,~u^{b}\pi^{a}_{~b} = 0$. Let us mention that all the curvature invariants for the metric (2.2) are finite (for example, the Ricci scalar is $-2b^{2}/r^{4}$ and the Kretschmann scalar is $12b^{4}/r^{8}$). With $b = l_{P}$, the energy density becomes $\rho = - \hbar c/r^{4}$. Hence, it no longer depends on the Newton constant $G$, having a pure quantum structure (similar for the pressures). In addition, its expression resembles the Casimir energy density between two planar conducting plates.

It is worth to see what type of field gives such a $T^{a}_{~b}$ as in (3.1). Harris \cite{EH} analysed in detail that issue and he found that a massless scalar field $\phi$ (whose Lagrangean density contains a negative kinetic term) fits perfectly with the problem at hand. His $T^{a}_{~b}$ is proportional to $1/r^{4}$ and is sourced by a charge $Q$ originating from the field $\phi(r)$, with $g^{ab}\nabla_{a}\phi \nabla_{b}\phi = 0$ and appearing as a constant of integration. A comparison with our situation gives $Q = b/\sqrt{8\pi G}$. We should also note that the flare-out condition \cite{SK} 
 \begin{equation}
 \frac{B - rB'}{2B^{2}} = \frac{r}{b^{2}} > 0
 \label{3.4}
 \end{equation} 
 is obeyed ($B' = dB/dr$).
 
 It is worth to notice that the traversable WH geometries are constrained by the sa-called Quantum Inequalities (QI) \cite{FR} that limit the magnitude and the spatial and temporal extent of negative energy rooted from the violation of the WEC. Ford and Roman \cite{FR} proved that the QI bound can be written as
  \begin{equation}
  \frac{\tau_{s}}{\pi} \int^{\infty}_{-\infty} \frac{< T_{ab}u^{a} u^{b}> d\tau }{\tau^{2} + \tau_{s}^{2}} \geq -\frac{3}{32\pi^{2}\tau_{s}^{4}},
 \label{3.5}
 \end{equation} 
where $\tau$ is the freely-falling observer proper time, $\tau_{s}$ is the sampling time and$ <T_{ab}u^{a}u^{b}>$ is the expectation value of the local energy density for a massless minimally-coupled scalar field. As we shall see in Sec.5, a static observer is geodesic for our metric (2.2). Therefore, the energy density seen by this observer is constant. For the spacetime under consideration, all the curvature components have their maximum value $1/b^{2}$ at the throat, the same being true for the energy-momentum tensor components (for example, $|\rho_{max}| = 1/8\pi b^{2}$ at the throat). Let us choose, following the authors of \cite{FR}, $\tau_{s} = r_{m}f = bf << r_{c}$ (with $f << 1$), where $r_{m} = b$ is the minimal length scale and $r_{c} \approx 1/\sqrt{R_{max}}$ is the smallest proper radius of curvature. One finds for our metric that $r_{m} = r_{c} = b$. Therefore, the constraint (3.5) gives us, for a static observer at $r = r_{0} = b$
  \begin{equation}
  -\frac{1}{8\pi b^{2}} \geq -\frac{3}{32\pi^{2}\tau_{s}^{4}},
 \label{3.6}
 \end{equation} 
 With $\tau_{s} = fr_{0}$ one obtains
    \begin{equation}
    b \leq \frac{l_{P}}{2f^{2}},~~~(\pi \approx 3)
 \label{3.7}
 \end{equation} 
But our choice for the magnitude of $b$ was of the order of the Planck length $l_{P}$ for microscopic WHs and so the relation (3.7) is obeyed when the FR prescription $f << 1$ is used. We arrived at the conclusion that the QI (3.5) is satisfied by our WH (2.2).

 \section{Energetic considerations}
 From the expressions (3.1) for the components of $T^{a}_{~b}$ it is clear that the energy conditions are not observed. The weak and dominant energy conditions require $\rho \geq 0$, the null and strong energy conditions impose $\rho + p_{i} \geq 0 (i = 1, 2, 3)$ and they are not satisfied for our exotic matter ($\rho < 0$ is necessary to hold the WH open). 
 
 We wish to compute now the Komar energy $W_{K}$ for the spacetime (2.2). We have \cite{TP, HC2}
  \begin{equation}
  W = 2 \int(T_{ab} - \frac{1}{2} g_{ab}T^{c}_{~c})u^{a} u^{b} N\sqrt{\gamma} d^{3}x.
 \label{4.1}
 \end{equation} 
 $N = \sqrt{-g_{00}}$ is the lapse function and $\gamma$ is the determinant of the spatial metric. With $T^{a}_{~a} = b^{2}/4\pi r^{4}$, (4.1) yields $W_{K} = 0$. In other words, the contribution from the energy density and pressures cancels out. In contrast, the $ADM$ energy is negative (only the energy density contributes)
    \begin{equation}
    W_{ADM} = 2\int{4\pi r^{2}\rho(r)dr} = \frac{b^{2}}{r}|^{\infty}_{b} = - b
 \label{4.2}
 \end{equation} 
 (the factor of 2 is rooted from the two WH sheets - see also \cite{PK}). We found that $W_{ADM}$ is (minus) the Planck energy . We may be also interested to write down the energy of the matter (the scalar field in our case), without that of the gravitational field \cite{SW} 
     \begin{equation}
    W_{sc} = 2\int{\sqrt{\gamma}\rho(r)drd\theta d\phi} = -b^{2}\int{\frac{dr}{r\sqrt{r^{2} - b^{2}}} = b~ arcsin\frac{b}{r}|^{\infty}_{b} = -\frac{\pi}{2}b}.
 \label{4.3}
 \end{equation} 
 This result has been previously obtained by Ellis \cite{HE} using different coordinates (his Eq. 64 for $E_{S}$). However, he assumed from the beginning that $E_{S} \geq 0$ and, therefore, our result matches his result up to a minus sign. 
 
 Let us compute now the quasilocal Misner - Sharp mass. It is given by \cite{CG1, NY, GGMP, HC1}
   \begin{equation}
  1 - \frac{2E_{MS}}{r} = g^{ab} \nabla_{a}r ~\nabla_{b}r.
 \label{4.4}
 \end{equation}
 One finds for the geometry (2.2) that
    \begin{equation}
  E_{MS}(r) = \frac{b^{2}}{2r}.
 \label{4.5}
 \end{equation}   
 A remark is in order here: if all fundamental constants are introduced in (4.5), one obtains $E_{MS} = \hbar c/2r$, i.e. it does not depend on $G$. This is a consequence of the $b^{2}$-factor on the r.h.s. of (4.5). We reached the same conclusion for the energy density $\rho$ from (3.1). As far as the Ricci part of $E_{MS}$ is concerned, we get \cite{HC3} 
 \begin{equation}
  E_{R} = \frac{4\pi}{3}r^{3}(\rho - p_{r} + p_{\theta}) = \frac{b^{2}}{6r}.
 \label{4.6}
 \end{equation}    
The Weyl part related to the energy of the gravitational field yields
 \begin{equation}
  E_{W} = E_{MS} - E_{R} = \frac{b^{2}}{3r}.
 \label{4.7}
 \end{equation}    
 
\section{Timelike geodesics}
We would first emphasize that the study of geodesics needs a differentiable manifold (a smooth spacetime). Nevertheless, we assumed our WH belongs to the Planck world (namely, its throat size is of the order of the Planck length $l_{P}$), where quantum fluctuations are supposed to exist and the spacetime smoothness seems to break down. However, we may adopt here the Di Casola et al. \cite{CLS} prescription and consider that a mesoscopic region may exist between the microscopic domain of quantum gravity and the macroscopic scales described by general relativity, where spacetime is still a differentiable pseudo-Riemannian manifold. The transition region covers the range between $l_{P}$ and some $\ell >> l_{P}$ (but still microscopic), where $\ell$ is the transition scale. Current observations imply that the scale $\ell$ must be much smaller than any curvature radius associated with macroscopic gravitational fields. We also have to remind that the constraint (3.5) works for a geodesic observer to whom one may also apply the Horowitz-Ross condition \cite{HR, NZK} $r_{0} > l_{P}/\sqrt{1-v_{0}^{2}}$, where $r_{0}$ is the throat radius and $v_{0}$ is the geodesic particle velocity at the throat. With $v_{0}$ close to the speed of light, we could obtain $r_{0} >> l_{P}$ such that the ''boosted'' $r_{0}$ enters the mesoscopic region, reaching the particle physics scale. In addition, Nandi et al. \cite{NZK} argued that an elementary particle could travel through the WH if its velocity is $v_{0} \approx 1$, even though its Bohr radius is much larger than $l_{P}$. 

Let us pass now to the geodesic equations. Since the spacetime (2.2) is static and using the standard procedure \cite{SF, CB, PK}, we have the two first integrals
 \begin{equation}
\dot{t} = \frac{dt}{d\lambda} = E,~~~~\dot{\phi} = \frac{d\phi}{d\lambda} = \frac{L}{r^{2}}
 \label{5.1}
 \end{equation}    
($t$ and $\phi$ are cyclic coordinates), where $\lambda$ is the affine parameter along the geodesics (the proper time for the timelike ones), $E$ is the energy per unit mass of the test particle and $L$ represents the angular momentum per unit mass. Using (5.1) for the metric (2.2) one obtains
 \begin{equation}
1 + \frac{\dot{r}^{2}}{1 - \frac{b^{2}}{r^{2}}} + \frac{L^{2}}{r^{2}} = E^{2},~~~(E \geq 1)
 \label{5.2}
 \end{equation}    
which can be written as 
 \begin{equation}
 \dot{r}^{2} + 1 + \frac{b^{2}(E^{2} - 1) + L^{2}}{r^{2}} - \frac{b^{2}L^{2}}{r^{4}} = E^{2}.
 \label{5.3}
 \end{equation}    
The last three terms on the l.h.s. of (5.3) plays the role of an effective potential and the geodesics could be clasified in terms of the values of the parameters $E, L$ and $b$. However, we are specially interested of the radial geodesics, namely $L = 0$. One obtains from (5.3)
 \begin{equation}
\left(\frac{dr}{dt}\right)^{2} + \frac{b^{2}(E^{2} - 1)}{E^{2}r^{2}} = 1 - \frac{1}{E^{2}}
 \label{5.4}
 \end{equation}    
which yields the hyperbolic trajectories
 \begin{equation}
r(t) = \sqrt{(1 - \frac{1}{E^{2}})t^{2} + b^{2}} = \sqrt{v_{\infty}^{2}t^{2} + b^{2}}
 \label{5.5}
 \end{equation}    
with the initial condition $r(0) = b = r_{min}$, $v_{\infty} = \sqrt{E^{2} - 1}/E$ and $t\in(-\infty, \infty)$. We notice that any static observer $r = const.$ is geodesic because $E = 1$ in that case. This is a consequence of the fact that $g_{00} = -1$ in Eq. (2.2). We may actually calculate the covariant acceleration 
 \begin{equation}
a^{b} = u^{a}\nabla_{a}u^{b}
 \label{5.6}
 \end{equation}    
of a static observer with $u^{b} = (1, 0, 0, 0)$ and obtain simply $a^{b} = 0$. The particle velocity from (5.5) is 
 \begin{equation}
\frac{dr}{dt} = \frac{(E^{2} - 1)t}{E\sqrt{(E^{2} - 1)t^{2} + b^{2}E^{2}}}
 \label{5.7}
 \end{equation}    
The velocity grows from $-v_{\infty}$ at $t = -\infty$ to zero at $t = 0$ and then to $v_{\infty}$ at $t = \infty$. The particular case $b = 0$ leads to $dr/dt = v_{\infty} = const.$ for all time, i.e. the inertial motion. With $b$ of the order of the Planck length, $dr/dt$ tends to $v_{\infty}$ very fast so that, macroscopically, the motion is uniform.
From (5.7) we can obtain
 \begin{equation}
\frac{d^{2}r}{dt^{2}} = \frac{b^{2}v_{\infty}^{2}}{(v_{\infty}^{2}t^{2} + b^{2})^{3/2}},
 \label{5.8}
 \end{equation}  
 with constant proper acceleration of magnitude $1/b$. Because $d^{2}r/dt^{2} > 0$ for any $t$, $dr/dt$ is monotonic so that it reaches its maximum value $v_{\infty}$ when $t \rightarrow \infty$.

As we remarked at the beginning of this section, our classical equation of motion (5.5) is not appropriate when quantum fluctuations are taken into account and the notion of classical trajectory looses its meaning. One means our geodesic calculations apply from the mesoscopic regime to macroscopic scales, viz. from $r > \ell$, or $t > \sqrt{\ell^{2} - b^{2}}/v_{\infty}$.

\section{Null geodesics}
Plugging $ds^{2} = 0$ in the metric (2.2) and using (5.1) one obtains
 \begin{equation}
\dot{r}^{2} + \frac{b^{2}E^{2} + L^{2}}{r^{2}} - \frac{b^{2}L^{2}}{r^{4}} = E^{2}.
 \label{6.1}
 \end{equation}    
or, for the angular trajectory
 \begin{equation}
\frac{d\phi}{dr} = \frac{d}{\sqrt{(r^{2} - b^{2})(r^{2} - d^{2})}}
 \label{6.2}
 \end{equation}    
where $d = L/E$ is the impact parameter. The null geodesics (6.1) were extensively studied by Ellis \cite{HE} in the coordinates (2.3) (see also \cite{CC, GC}). Therefore, we only intend to analyse the null radial geodesics in the geometry (2.2). With $L = 0$ in (6.1) we have 
 \begin{equation}
\frac{dr}{dt} = \pm \sqrt{1 - \frac{b^{2}}{r^{2}}},~~~(r > b)
 \label{6.3}
 \end{equation}    
which yields
 \begin{equation}
r(t) = \sqrt{t^{2} + b^{2}},
 \label{6.4}
 \end{equation}    
with $r(0) = b$. For negative $t$, $r$ decreases from $\infty$ to $b$ at $t = 0$ in the lower universe and then reaches again $\infty$ for $t > 0$, in the upper universe. When (6.4) is written as 
 \begin{equation}
r^{2} - t^{2} = b^{2},
 \label{6.5}
 \end{equation}    
one sees that the null radial equation of motion is Lorentz invariant. That remind us the bubble of Coleman and de Luccia \cite{CL} whose surface traces out a hyperboloid with an O(3,1) invariance evolution. Almost immediately after its materialisation, their bubble accelerates to the speed of light because its initial radius is typically a quantity of subnuclear magnitude. Since we have chosen $b$ to be of the order of the Planck length, Eq. (6.5) introduces in fact a minimal Minkowski interval (see also \cite{KP}): $\sqrt{(\eta_{ab}x^{a}x^{b})_{min}} = b$. 

If the spacetime foam hypothesis proves to be valid then the Minkowski light cone is only an approximation valid macroscopically (when the Planck length is neglected). It should be noticed that in the coordinates (2.3) the radial null geodesics are straightlines, as in Minkowski space. However, the $l = 0$ surfaces have no zero area, but $4\pi b^{2}$.

\section{Dynamic wormhole}
 Although at macroscopic scales the spacetime appears smooth and simply connected, on Planck length scales it fluctuates quantum-mechanically, developing all kinds of topological structures, including WHs. A microscopic WH may be extracted from the foam to give birth to a macroscopic traversable WH. Redmount and Suen (RS) \cite{RS} see Lorentzian spacetime filled with many microscopic WHs, living for microscopic time periods and then pinching off. They found those WHs are quantum-mechanically unstable, like a classical stable black hole which however undergoes quantum Hawking evaporation. RS constructed a spherically-symmetric ''Minkowski wormhole'' by excising a sphere of radius $r = R(t)$ ($t$ - the Minkowski time coordinate) from two copies of the Minkowski space, identifying the two boundary surfaces $r = R(t)$. To obey Einstein's equations a surface stress tensor on the boundary $\Sigma$ was introduced. Outside the boundary both exterior spacetimes are flat. The boundary plays the role of the WH throat and the Einstein equations are equivalent with the Lanczos equations \cite{MV} 
  \begin{equation}
  -8\pi S^{i}_{~j} = \left[K^{i}_{~j} - \delta^{i}_{~j}K^{n}_{~n}\right]
 \label{7.1}
 \end{equation}    
with $S^{i}_{~j}$ the surface stress tensor (here $i, j = 0, 2, 3$), $K^{n}_{~n}$ - the trace of the extrinsic curvature of the boundary $\Sigma$ and $[...]$ stands for the jump of $K^{i}_{~j}$ when the boundary is crossed. 

 Let us find now the the extrinsic curvature tensor of the surface $F = r - R(t)$. It is given by \cite{LL}
  \begin{equation}
  K_{ij} = \frac{\partial x^{a}}{\partial \xi^{i}}\frac{\partial x^{b}}{\partial \xi^{j}}\nabla_{a}n_{b} = -n_{c}\left(\frac{\partial^{2}x^{c}}{\partial \xi^{i}\partial \xi^{j}} + \Gamma_{ab}^{c} \frac{\partial x^{a}}{\partial \xi^{i}}\frac{\partial x^{b}}{\partial \xi^{j}}\right)
 \label{7.2}
 \end{equation}      
where $\xi^{i}$ are the coordinates on $\Sigma$, the $\Gamma's$ are the affine connections in Minkowski space and $n_{c}$ is the unit normal to $\Sigma$. The spacetime metric is Minkowskian and the geometry on $\Sigma$ can be written as
  \begin{equation}
ds^{2}_{\Sigma} = - d\tau^{2} + R^{2}d\Omega^{2}
 \label{7.3}
 \end{equation}   
where $d\tau = \sqrt{1 - \dot{R}^{2}}dt$, $\tau$ is the proper time on $\Sigma$ and $\dot{R} = dR/dt$. The velocity 4-vector is 
  \begin{equation}
  u^{b} = \frac{dx^{b}}{d\tau} = \left(\frac{dt}{d\tau}, \frac{dR}{d\tau}, 0, 0\right)
 \label{7.4}
 \end{equation}   
with $u^{b}u_{b} = -1$. The unit normal to $\Sigma$ may be found from (7.4) and the relations $n^{b}n_{b} = 1$ and $n^{b}u_{b} = 0$. It is also given in general by
  \begin{equation}
  n_{a} = \frac{F_{,a}}{\sqrt{g^{ab}F_{,a}F_{,b}}}
 \label{7.5}
 \end{equation}   
where $F_{,a} = \partial F/\partial x^{a}$. The velocity $u^{b}$ from (7.4) yields
  \begin{equation}
  u^{b} = \left(\frac{1}{\sqrt{1 - \dot{R}^{2}}}, \frac{\dot{R}}{\sqrt{1 - \dot{R}^{2}}}, 0, 0\right)
 \label{7.6}
 \end{equation}   
whence
  \begin{equation}
  n_{b} = \left(\frac{\dot{R}}{\sqrt{1 - \dot{R}^{2}}}, \frac{-1}{\sqrt{1 - \dot{R}^{2}}}, 0, 0\right)
 \label{7.7}
 \end{equation}   
Eq. (7.2) gives the following components of the second fundamental form
  \begin{equation}
  K_{\tau \tau} = \frac{-\ddot{R}}{(1 - \dot{R}^{2})^{3/2}}, ~~~K_{\theta \theta} = \frac{R}{\sqrt{1 - \dot{R}^{2}}} = \frac{K_{\phi \phi}}{sin^{2}\theta},
 \label{7.8}
 \end{equation}   
with the trace
  \begin{equation}
  K^{i}_{~i} = \frac{\ddot{R}}{(1 - \dot{R}^{2})^{3/2}} + \frac{2}{R \sqrt{1 - \dot{R}^{2}}}.
 \label{7.9}
 \end{equation}   
Using (7.8) and (7.9), Eqs. (7.1) appear as
   \begin{equation}
  S_{\tau \tau} = -\frac{1}{2\pi R\sqrt{1 - \dot{R}^{2}}}, ~~~S_{\theta \theta} = \frac{R}{4\pi \sqrt{1 - \dot{R}^{2}}} +  \frac{R^{2}\ddot{R}}{4\pi (1 - \dot{R}^{2})^{3/2}} = \frac{S_{\phi \phi}}{sin^{2}\theta},
 \label{7.10}
 \end{equation}   
i.e. similar with the expressions obtained in \cite{RS}. Supposing that $S_{ij}$ on the throat corresponds to a perfect fluid
   \begin{equation}
   S_{ij} = (p_{s} + \sigma)u_{i}u_{j} + p_{s}h_{ij}
 \label{7.11}
 \end{equation}   
 where $h_{ij} = (-1, R^{2}, R^{2}sin^{2}\theta)$ is the metric on the boundary, we have $\sigma = S_{\tau \tau}$ for the surface energy density and $p_{s} = S_{\theta \theta}/R^{2}$ for the surface pressure. 
 
 To find the equation of motion for the throat, we need now an equation of state relating $\sigma$ and $p_{s}$. RS chose $\sigma = -4p_{s}$ as equation of state but in this case the action integral (whence the equation of motion was obtained) has a complicated ''kinetic term''. Our choice for the equation of state is simply $p_{s} = -\sigma$, as for a domain wall \cite{IS} because one seems to be the most appropriate conjecture for the exotic matter characterising a WH. That choice leads to the equation of motion 
   \begin{equation}
   \frac{\ddot{R}}{1 - \dot{R}^{2}} - \frac{1}{R} = 0, ~~~\dot{R} \neq 1,
 \label{7.12}
 \end{equation} 
 or
    \begin{equation}
    R \ddot{R} + \dot{R}^{2} - 1 = 0,
 \label{7.13}
 \end{equation}   
 which has the solution
    \begin{equation}
    R(t) = \sqrt{t^{2} + b^{2}}.
 \label{7.14}
 \end{equation}   
using appropriate initial conditions. An action corresponding to (7.13) appears as
    \begin{equation}
    S = \int{\frac{b^{2}}{R}\sqrt{1 - \dot{R}^{2}}dt }
 \label{7.15}
 \end{equation}   
whence the Lagrangean is given by (not to be confused with the angular momentum $L$ from Sec.5 and 6)
    \begin{equation}
   \textit{L} = \frac{b^{2}}{R}\sqrt{1 - \dot{R}^{2}}
 \label{7.16}
 \end{equation}   
(the factor $b^{2}$ is necessary so $\textit{L}$ has units of length). The canonical momentum will be
    \begin{equation}
  p = \frac{\partial \textit{L}}{\partial \dot{R}} = -\frac{b^{2}\dot{R}}{R \sqrt{1 - \dot{R}^{2}}},
 \label{7.17}
 \end{equation}   
which yields the Hamiltonian
    \begin{equation}
  H = p \dot{R} - \textit{L} = -\frac{b^{2}}{R \sqrt{1 - \dot{R}^{2}}}.
 \label{7.18}
 \end{equation}   
To find the direct relation between $p$ and $H$ we get rid of $\dot{R}$ from the last two equations to obtain
    \begin{equation}
  H = -\sqrt{p^{2} + \left(\frac{b^{2}}{R}\right)^{2}}.
 \label{7.19}
 \end{equation}  
We see that $-b^{2}/R = -\hbar/cR$ plays the role of a negative mass $M$ of the ''particle'' (expanding WH throat in our case), namely $M = -\hbar/cR$. So $R = \hbar/|M|c$ appears as being the Compton wavelength associated to the mass $M$. For $t >> b$, $R(t) \approx t$ so that $|M|c^{2}t = \hbar$, which looks like an uncertainty relation. One could also see from (7.18) that $H$ has the form of a Lorentz-boosted negative mass. This simple WH geometry seems to represent a spacetime foam structure unstable against growth to macroscopic size \cite{RS}. Thus, the behavior of the throat resembles that of a particle initially confined to a well or of an $\alpha$-particle in a heavy nucleus.

When (7.14) is used in the expression for the WH energy (7.18), we get $H = -b$, i.e. the ADM mass (4.2) for the Morris-Thorne wormhole. Moreover, the action $S$ from (7.15) acquires the form
    \begin{equation}
    S = \int{\frac{b^{3}}{t^{2} + b^{2}}dt} = b^{2} arctan\frac{t}{b}
 \label{7.20}
 \end{equation}   
which is bounded ($\pi \hbar/2 > S > -\pi \hbar/2$).

Using (7.10), the expression (7.14) for $R(t)$ yields
    \begin{equation}
    \sigma = -p_{s} = -\frac{1}{2\pi b}
 \label{7.21}
 \end{equation}   
 We obtained, indeed, $\sigma < 0$, as it should be for exotic matter.
 
 In terms of the proper time on $\Sigma$, (7.14) can be written as $R(\tau) = b~cosh(\tau/b)$. Therefore, the geometry on the throat appears as
   \begin{equation}
ds^{2}_{\Sigma} = - d\tau^{2} + b^{2}cosh^{2}\frac{\tau}{b}d\Omega^{2},
 \label{7.22}
 \end{equation} 
 which is the closed de Sitter space in three dimensions. From (7.14) we have 
 \begin{equation}
 1 - \dot{R}^{2} = \frac{b^{2}}{R^{2}}, ~~~\ddot{R} = \frac{b^{2}}{R^{3}}.
 \label{7.23}
 \end{equation}
 Therefore, the component of the acceleration of the throat, normal to $\Sigma$ will be \cite{MV}
  \begin{equation}
  A_{\bot} \equiv n_{b}A^{b} = -K_{\tau \tau} = \frac{1}{b} = -2\pi \sigma > 0,
 \label{7.24}
 \end{equation}
 where $A^{b}$ is built with $u^{b}$ from (7.6). So we obtained the same evolution of the WH throat as Ipser and Sikivie for their domain wall which in Minkowskian coordinates is not a plane at all but rather an accelerating sphere, expanding with the acceleration $2\pi |\sigma|$.
 
 A remark is in order here. The radial null geodesics (6.4) in the static WH geometry (2.2) are similar with the equation of motion (7.14) of the dynamic WH throat. Note that the spacetime (2.2) is curved and the region $r < b$ is absent from the manifold. We identify the two processes and assume that actually the null particles are carried by the WH throat during their propagation (see \cite{HC4}). In other words, the throat plays the role of a de Broglie pilot wave, dragging the null particles with it.
 
 \section{Conclusions}
 We investigated in this paper a special case of the Misner - Thorne Lorenzian traversable WH. We showed that the spacetime corresponding to the two WH sheets is endowed with an anisotropic fluid with negative energy density $\rho$ and radial pressure $p_{r}$. The WH throat radius was taken of the order of the Planck length and so $\rho$ and $p_{r}$ do not depend on the Newton constant but only on $\hbar$ and $c$, having a Casimir type dependence on $r$. While the Komar energy of our wormhole is vanishing, we obtained that the ADM energy equals (minus) the Planck energy. Macroscopically, the timelike geodesics are straightlines but they become hyperbolae microscopically, when the Planck length is taken into account. A similar effect appeared for null geodesics. The null trajectories are hyperbolae too but after large time period they become straightlines, moving asymptotically with the velocity $c = 1$. We also conjectured a minimum Minkowski interval, in the spirit of Kothawala and Padmanabhan. Using a different equation of state compared to that of Redmount and Suen, we found that a dynamic WH expands hyperbolically in a fashion similar with the Coleman and de Luccia bubble or Ipser and Sikivie domain wall, i.e. a Lorentz-invariant expansion.
 
 We also notice the role played by WHs in Einstein-Podolsky-Rosen (EPR) pair and Einstein-Rosen (ER) bridge \cite{MS, JK}. Maldacena and Susskind showed that EPR pairs and non-traversable WHs are equivalent description of the same physics.In addition, in the holographic dual the quantum entanglement is encoded in a geometry of a non-traversable WH on the worldsheet of the flux tube connecting the EPR pair. It was pointed out \cite{JS, CGP} that the bulk dual of an entangled pair of a quark and anti-quark corresponds to the Lorenzian continuation of the tunneling instanton describing Schwinger pair creation in a strong electric field.
 
 \textbf{Acknowledgements}\\
 I would like to thank one of the anonymous referees for useful comments and suggestions which led to a substantially improved manuscript.

\end{document}